%% file: main.tex
\documentclass[aps,prl,reprint,superscriptaddress]{revtex4-1}

\input{packages}

\input{definitions}

\begin{document}

\title{In situ thermometry of a cold Fermi gas via dephasing impurities}

\author{Mark T. Mitchison}
\email{mark.mitchison@tcd.ie}
\affiliation{School of Physics, Trinity College Dublin, College Green, Dublin 2, Ireland}

\author{Thom\'{a}s Fogarty}
\affiliation{Quantum Systems Unit, Okinawa Institute of Science and Technology Graduate University, Onna, Okinawa 904-0495, Japan}

\author{Giacomo Guarnieri}
\affiliation{School of Physics, Trinity College Dublin, College Green, Dublin 2, Ireland}

\author{Steve Campbell}
\affiliation{School of Physics, University College Dublin, Belfield Dublin 4, Ireland}

\author{Thomas Busch}
\affiliation{Quantum Systems Unit, Okinawa Institute of Science and Technology Graduate University, Onna, Okinawa 904-0495, Japan}

\author{John Goold}
\email{gooldj@tcd.ie}
\affiliation{School of Physics, Trinity College Dublin, College Green, Dublin 2, Ireland}

\date{\today}

\begin{abstract}
The precise measurement of low temperatures is a challenging, important and fundamental task for quantum science. In particular, in situ thermometry is highly desirable for cold atomic systems due to their potential for quantum simulation. Here we demonstrate that the temperature of a non-interacting Fermi gas can be accurately inferred from the non-equilibrium dynamics of impurities immersed within it, using an interferometric protocol and established experimental methods. Adopting tools from the theory of quantum parameter estimation, we show that our proposed scheme achieves optimal precision in the relevant temperature regime for degenerate Fermi gases in current experiments. We also discover an intriguing trade-off between measurement time and thermometric precision that is controlled by the impurity-gas coupling, with weak coupling leading to the greatest sensitivities. This is explained as a consequence of the slow decoherence associated with the onset of the Anderson orthogonality catastrophe, which dominates the gas dynamics following its local interaction with the immersed impurity.
\end{abstract}

\maketitle

Temperature measurements are crucial for many experiments using ultracold atomic gases, for example when calibrating quantum simulators~\cite{Hofstetter_2018,Tarruell_2018} or when determining equations of state~\cite{Navon729,Ku563}. Unfortunately, standard thermometry techniques such as time-of-flight or in situ absorption imaging are inherently destructive and involve integration over the line of sight~\cite{Inguscio:2007cma}. A minimally disturbing method to probe \textit{local} temperature profiles would be beneficial for numerous experimental scenarios of current interest, including thermalisation dynamics after a quench~\cite{Sadler_2006,Hofferberth2007,Trotzky2012,Gring2012,Cheneau2012} or energy transport between separate thermal reservoirs~\cite{Brantut_2013,Krinner2017}. Further motivation is provided by recent progress in the preparation of homogeneous ultracold gases~\cite{Gaunt2013,Schmidutz2014,Navon2015,Mukherjee2017,Hueck2018,Mukherjee2019,Yan2019}, whose constant density distribution does not carry information on temperature, thus rendering standard in situ thermometry techniques ineffective. 

An appealing alternative method of in situ thermometry exploits impurity atoms as probes embedded within the ultracold gas~\cite{Roati2002, Silber2005,Wenz2013,Massignan2014,Schmidt2018}. The advantage of this approach is that a single atom can be confined to sub-micron length scales and its state is relatively easy to characterise. For example, temperature can be inferred by allowing the impurities to equilibrate with the gas and then measuring their mean energy or a similar observable~\cite{Hangleiter2015,Correa2015,Mehboudi2019}. This method has proved useful in several recent experiments~\cite{Regal2005,Spiegelhalder2009,Nascimbene2010,McKay2010,Olf2015,Hohmann2016,Lous2017} but becomes challenging at low temperatures where equilibration is slow and the probe's energy levels must be finely tuned~\cite{Correa2015,Paris2015,Pasquale2016,Hovhannisyan2018,CampbellQST,Potts2019,Joergensen2020}. These limitations can be overcome by harnessing the probe's \textit{non-equilibrium} dynamics for thermometry~\cite{Ferrari1999,Hangleiter2015, Jevtic2015,Mancino2017,Hofer2017,Cavina2018,Bouton2020}. Perhaps the most extreme example is pure dephasing, where the energy of the probe is conserved and thus normal thermalisation is completely suppressed. Nevertheless, coherences between the probe energy eigenstates can develop into correlations with the environment that are sensitive to temperature~\cite{Bruderer_2006,Sabin2014,Johnson2016,Razavian2019}.
    
In this letter, we apply this idea to address a long-standing challenge in cold-atom physics: namely, thermometry of degenerate Fermi gases~\cite{McKay2011,Onofrio_2016}. Specifically, we propose to measure the temperature of an ultracold Fermi gas by observing the non-equilibrium dephasing dynamics of impurities immersed within it. We focus on a promising setup that has already been realised in the laboratory~\cite{Cetina2015,Cetina2016,Ness2020}, where the gas atoms effectively interact only with the impurities and not with each other. In this setting, the Anderson orthogonality catastrophe (OC)~\cite{Anderson1967,Fogarty2020} imprints characteristic signatures on the decoherence dynamics of the impurity~\cite{Sindona2013,Sindona2014,Cosco2018,Tonielli2019}, which can be observed using Ramsey interferometry~\cite{Goold2011,Knap2012,Cetina2016}. The optimal precision of our thermometry protocol can be evaluated in terms of the quantum Fisher information, and we reveal a tradeoff between measurement time and precision controlled by the impurity-gas interaction strength. Since this coupling can be experimentally tuned over a wide range of values by means of Feshbach resonances~\cite{Cheng2010}, our approach allows for precise in situ thermometry of homogeneous Fermi gases in the deeply degenerate regime. 

\textit{Thermometry by qubit dephasing.---}Let us begin with the general scenario of a two-level probe (qubit) $S$ undergoing pure dephasing induced by its environment $E$. The total Hamiltonian is $\hat{H}= \hat{H}_S + \hat{H}_E + \hat{H}_I$, where $\hat{H}_I$ is an interaction which satisfies $[\hat{H}_S,\hat{H}_I] = 0$. We assume that the system is initially prepared in the product state $\hat{\rho} = \ket{+}\bra{+} \otimes \hat{\rho}_E(T)$, where $\hat{\rho}_E(T)$ is a thermal state of the environment at temperature $T$ and $\ket{+} = (\ket{0}+\ket{1})/\sqrt{2}$ is an equal superposition of the qubit's energy eigenstates. The populations of these eigenstates are strictly conserved in time, while the qubit coherences decay according to the decoherence function
\begin{equation}\label{decoherence_function}
    v(t)= \Tr_E \left[\ee^{\ii\hat{H}_{1}t/\hbar}\ee^{-\ii\hat{H}_{0}t/\hbar}\hat{\rho}_{E}(T)\right],
\end{equation}
where $\hat{H}_j = \bra{j}\hat{H}_E + \hat{H}_I\ket{j}$ is the Hamiltonian of the environment conditioned on the qubit eigenstate $j = 0,1$. In a frame rotating at the qubit precession frequency, the state of the qubit is given by $\hat{\rho}_S = \tfrac{1}{2}(1 + \mathbf{v}\cdot \bm{\sg})$, where $\mathbf{v} = (\Re [v], \Im [v], 0)$ is the Bloch vector and $\bm{\sg} = (\sg_x,\sg_y,\sg_z)$ are Pauli matrices.

The initial temperature of the gas parametrises the probe state $\hat{\rho}_S(T)$ via the decoherence function in Eq.~\eqref{decoherence_function}. If the dependence of $v(t)$ on $T$ is well understood, this temperature can therefore be inferred from the statistics of measurements made on a large ensemble of identically prepared probes. Any such temperature estimate carries an unavoidable uncertainty due to the random character of quantum measurement and the finite size of the ensemble. To find the optimal measurement that minimises this uncertainty, we appeal to the theory of quantum parameter estimation~\cite{Paris2009,Toth2014,Mehboudi2019a}. 

In general, a measurement is described by a positive operator-valued measure (POVM) $\{\hat{\Pi}(\xi)\}$ satisfying $\int\dd\xi\,\hat{\Pi}(\xi) = \mathbbm{1}$, where $\xi$ labels the possible outcomes. Performing $N$ independent measurements on identical qubit preparations yields the random outcomes $\bm{\xi}=\{\xi_{1},\xi_{2},\dots,\xi_{N}\}$, from which a temperature prediction is generated via an estimator function $T_{\rm est}(\bm{\xi})$. We consider unbiased estimators with $\mathbb{E}[T_\mathrm{est}] = T$, where 
\begin{equation}
    \label{classical_average}
    \mathbb{E}\left[T_\mathrm{est}\right] = \int\dd\xi_1 \cdots \int\dd \xi_N\, p(\xi_1|T) \cdots p(\xi_N|T) T_\mathrm{est}(\bm{\xi}),
\end{equation}
and $p(\xi|T) = \Tr[\hat{\Pi}(\xi)\hat{\rho}_S(T)]$. The expected uncertainty of the temperature estimate is quantified by $\Delta T^2 = \mathbb{E}[\left(T_{\rm est}-T\right)^2]$ and the error of any unbiased estimator obeys the quantum Cram\'er-Rao bound $\Delta T^2 \geq 1/N\mathcal{F}_T \geq 1/N\mathcal{F}^Q_T$~\cite{Braunstein1994}. Here $\mathcal{F}_T$ is the Fisher information associated with the measurement,
\begin{equation}
    \label{Fisher_information}
\mathcal{F}_T = \int \dd \xi \, p(\xi|T) \left( \frac{\partial \ln p(\xi|T)}{\partial T}\right)^2 = \frac{1}{\langle \Delta \hat{X}^2 \rangle }\left(\frac{\partial \langle \hat{X} \rangle}{\partial T}\right)^2,
\end{equation}
and the second equality holds for projective measurements on a two-level system, with $\langle \hat{X}\rangle$ and $\langle \Delta \hat{X}^2 \rangle$ the mean and variance of the measured observable $\hat{X}$. The Fisher information of any POVM is bounded by the quantum Fisher information (QFI) $\mathcal{F}^Q_T = \max_{\hat{X}}\mathcal{F}_T(\hat{X})= \mathcal{F}_T(
\hat{\Lambda}_T)$ and the maximum is achieved by projective measurements of a specific observable: the symmetric logarithmic derivative (SLD), denoted by $\hat{\Lambda}_T$~\cite{Paris2009}. We also define the quantum signal-to-noise ratio (QSNR) $\mathcal{Q}^2 = T^2\mathcal{F}_T^Q$, which bounds the signal-to-noise ratio as $T/\Delta T \leq \sqrt{N}\mathcal{Q}$. Hence, $\mathcal{Q}$ quantifies the ultimate sensitivity limit of our impurity thermometer.

For a qubit probe, the QFI has a simple expression in terms of the Bloch vector~\cite{Zhong2013}, and for pure dephasing it can be conveniently written in polar coordinates using $v = |v|\ee^{\ii\phi}$ as
\begin{equation}
    \label{QFI_qubit}
    \mathcal{F}_T^Q  = \frac{1}{1-|v|^2}\left(\frac{\partial |v|}{\partial T}\right)^2 + |v|^2 \left(\frac{\partial \phi}{\partial T}\right)^2  = \mathcal{F}^\parallel_T + \mathcal{F}^\perp_T .
\end{equation}
The QFI comprises two terms, respectively corresponding to the Fisher information for measurements of $\sg_\parallel = \cos(\phi) \sg_x + \sin(\phi) \sg_y$ and $\sg_\perp = \cos(\phi)\sg_y - \sin(\phi)\sg_x$, i.e.~parallel and perpendicular to the Bloch vector of $\hat{\rho}_S(T)$. Neglecting irrelevant shift and scale factors, the SLD is given by 
\begin{align}
    \label{SLD}
    \hat{\Lambda}_T & \propto \cos (\varphi)\sg_\parallel + \sin (\varphi) \sg_\perp, \quad \tan (\varphi) =\frac{ |v|(1-|v|)^2 \partial_T \phi}{\partial_T|v|}.
\end{align}
Since the SLD is optimal in the sense of the quantum Cram\'er-Rao bound, measuring $\hat{\Lambda}_T$ minimises the uncertainty in the temperature estimate due to the finite number of samples. Note that the SLD is temperature-dependent and thus some prior information on $T$ is assumed. In practice, measuring $\hat{\Lambda}_T$ requires an efficient prescription to evaluate $|v|$, $\phi$ and their temperature derivatives from an accurate theoretical model for $\hat{\rho}_S(T)$, as well as the ability to measure an arbitrary combination of $\hat{\sigma}_x$ and $\hat{\sigma}_y$.

\begin{figure}
    \centering
    \includegraphics[width=\linewidth]{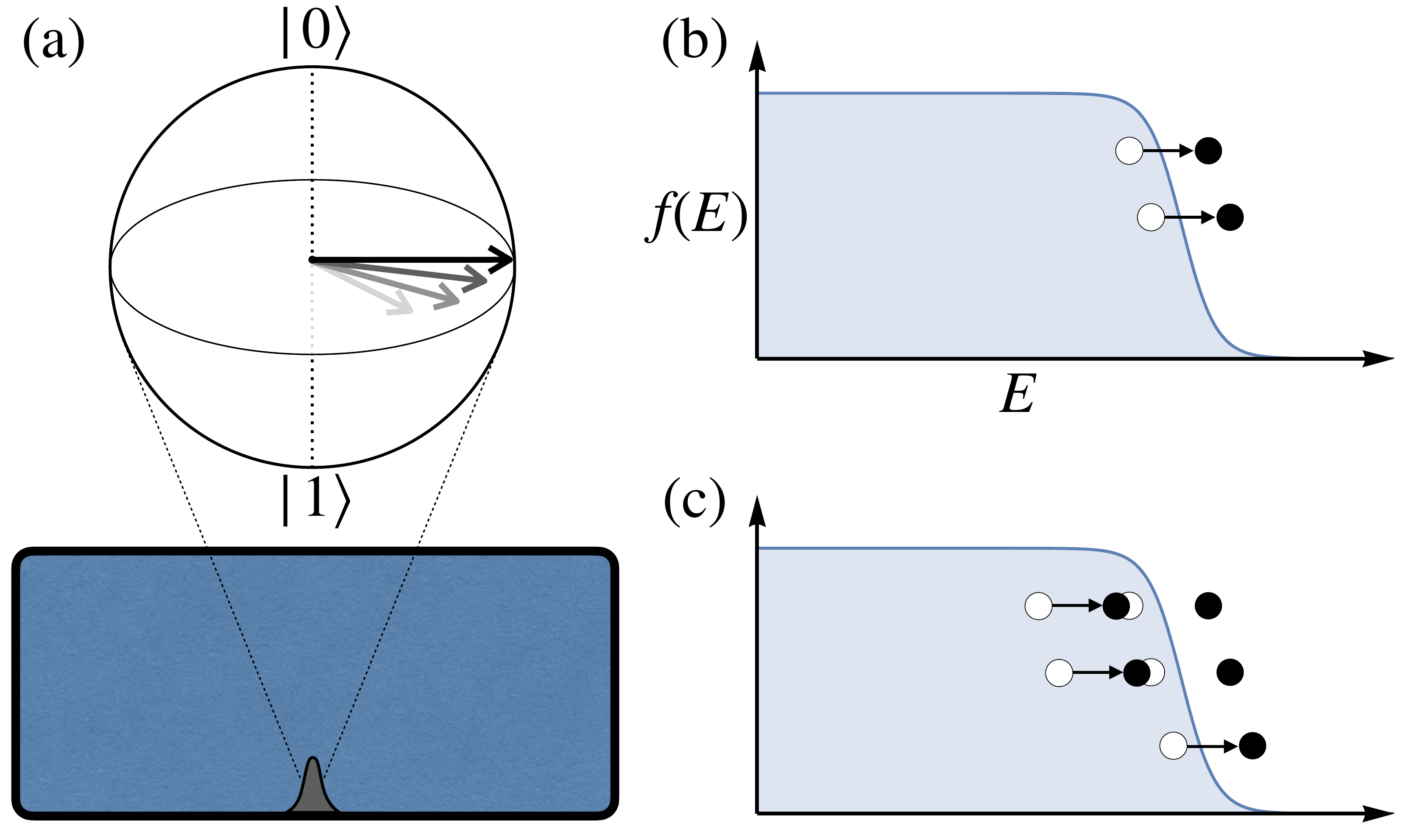}
    \caption{Schematic depiction of the system. (a)~A cold Fermi gas (blue) is perturbed by a localised impurity (grey) with two internal states that undergo pure dephasing. (b)~Scattering from the impurity disturbs the atoms' initial equilibrium distribution, $f(E)$. Pauli blocking restricts the resulting particle-hole excitations to a region near the Fermi surface. (c)~The creation of holes eventually allows further scattering to generate excitations deep within the Fermi sea.}
    \label{fig:schematic}
\end{figure}

\textit{Physical model.---} From here on, we focus on a scenario realised in recent experiments~\cite{Cetina2016}, which satisfies the aforementioned desiderata for optimal thermometry. Here, the qubit comprises two spin states of an impurity immersed in a spin-polarised Fermi gas (see Fig.~\ref{fig:schematic}). We assume that the impurity is confined to the ground state of a species-selective potential so that its kinetic energy can be neglected. The only relevant collision process at low temperatures is $s$-wave scattering, which does not occur between identical fermions due to wavefunction anti-symmetry. Therefore, the gas atoms do not interact with each other, while their coupling to the impurity is controlled by a spin-dependent $s$-wave scattering length. We assume that the impurity and the gas interact only when the impurity is in state $\ket{1}$, which can be achieved by tuning the scattering length for state $\ket{0}$ to zero via a Feshbach resonance~\cite{Cheng2010}. 

We consider the following interferometric protocol. The gas is prepared in a thermal state with the impurity in the non-interacting state $\ket{0}$, leading to an initial density matrix $\hat{\rho} = \ket{0} \bra{0} \otimes \hat{\rho}_E(T)$. A $\pi/2$-pulse then prepares the superposition state $\ket{0}\to\ket{+}$ and the system freely evolves for a time $t$, after which the qubit coherences are given by Eq.~\eqref{decoherence_function}. Finally, a second $\pi/2$-pulse is applied with a phase $\theta$ relative to the initial pulse and  the qubit's energy is projectively measured, giving a result proportional to $\cos(\theta)\langle\sg_x\rangle + \sin(\theta)\langle\sg_y\rangle$ on average. Repeating this procedure $N$ times --- or using $N$ independent impurities interacting with a single copy of the gas --- yields the expectation value of any combination of $\sg_x$ and $\sg_y$, e.g.~choosing $\theta = \phi + \varphi$ realises a measurement of $\hat{\Lambda}_T$.

For a non-interacting gas, the decoherence function can be computed \textit{exactly} using the Levitov formula~\cite{LevitovForm1, LevitovForm2} 
\begin{equation}
    \label{functional_determinant}
    v(t) = \det \left[1 - \hat{n} + \hat{n}\ee^{\ii\hat{h}_0t/\hbar}\ee^{-\ii\hat{h}_1t/\hbar}\right ],
\end{equation}
where $\hat{h}_1$ and $\hat{h}_0$ are single-particle Hamiltonians describing atoms in the gas with or without the impurity present, respectively. The initial thermal distribution is described by $\hat{n} = (\ee^{\beta(\hat{h}_0 - \mu)} + 1)^{-1}$, where $\beta = 1/k_B T$ and $\mu$ is the chemical potential. In general, we have
\begin{align}
    \label{single_particle_h0}
    \hat{h}_0 &= \frac{-\hbar^2}{2m}\nabla^2 + V_{\rm ext}(\rr), \\
    \label{single_particle_h1}
    \hat{h}_1 &= \hat{h}_0 + V_{\rm imp}(\rr),
\end{align}
where $m$ is the atomic mass, $V_{\rm ext}(\rr)$ is an external potential, $V_{\rm imp}(\rr) = \int\dd\rr' V_{\rm int}(\rr-\rr')|\chi(\rr')|^2$ is the scattering potential generated by a static impurity with wavefunction $\chi(\rr)$, and $V_{\rm int}(\rr)$ is the interatomic interaction potential. Collisions in the $s$-wave channel are described by the regularised pseudopotential $V_{\rm int}(\rr) = (2\pi \hbar^2 a/m_{\rm red}) \delta(\rr) (\partial/\partial r) r$, with $a$ the scattering length and $m_{\rm red}$ the reduced mass~\cite{Busch1998}. Crucially, Eq.~\eqref{functional_determinant} replaces a complex many-body expectation value with a determinant over single-particle states, allowing efficient computation of a temperature estimate from the experimental data.

\begin{figure}
    \centering
    \includegraphics[width=\linewidth, trim = 1cm 8cm 1cm 9.8cm, clip]{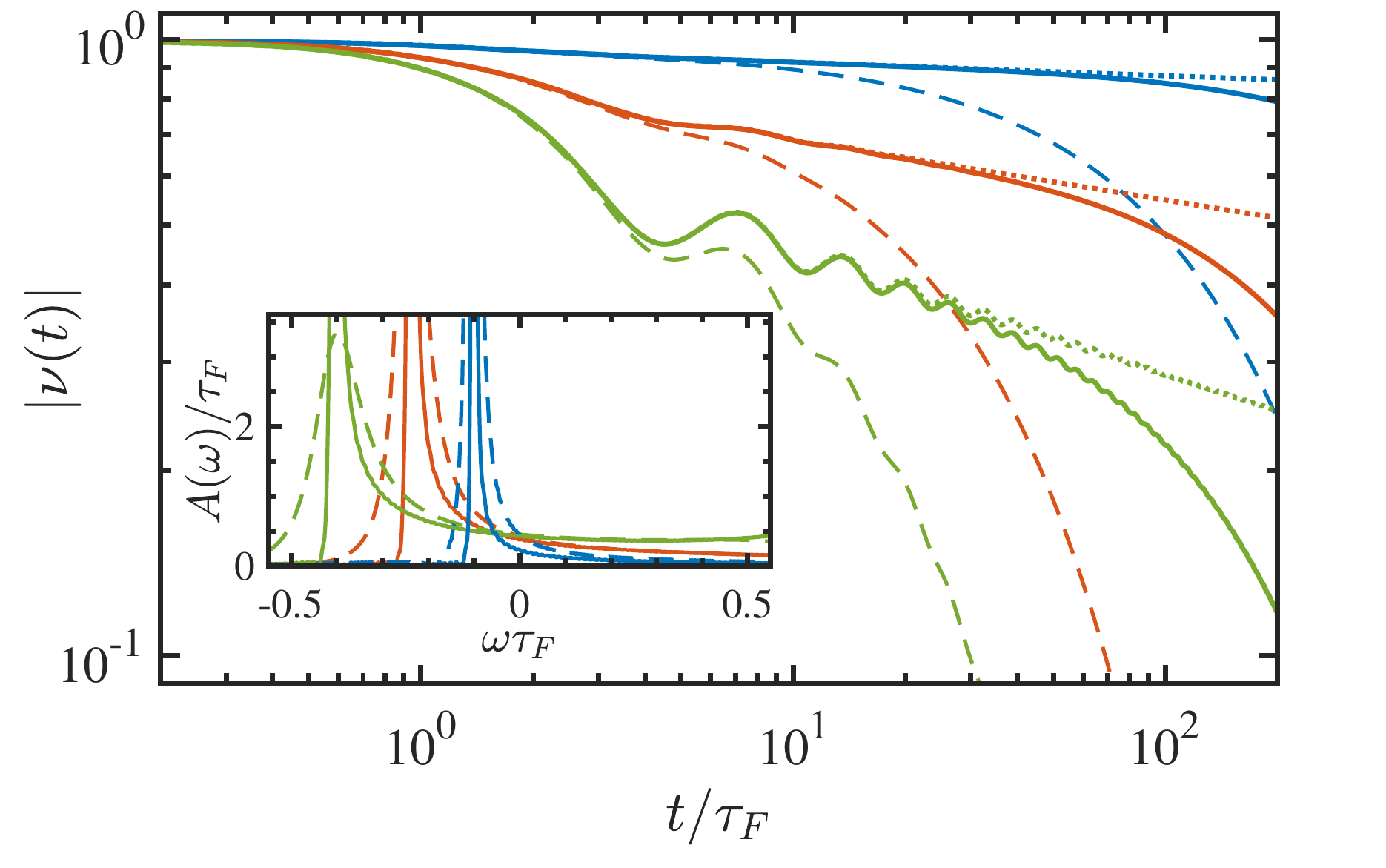}
    \caption{Decoherence functions (main) and absorption spectra (inset) for the homogeneous gas, with coupling $k_F a = -0.5$ (blue), $k_Fa = -1.5$ (red) and $k_F a = -6$ (green); and temperature $T = 0$ (dotted), $T=0.01T_F$ (solid) and $T=0.1T_F$ (dashed). Spectra for $T=0$ not shown.}
    \label{fig:vee}
\end{figure}

\textit{Decoherence in a homogeneous gas.---}From here on, we focus on a three-dimensional (3D), homogeneous Fermi gas ($V_{\rm ext} = 0$) of mean density $\bar{n}$ that is trapped in a box large enough to prevent finite-size effects. We assume the impurity is tightly confined so that the infinite-mass approximation is valid, i.e.\ $|\chi(\rr)|^2 \approx \delta(\rr)$ and $m_\mathrm{red} = m$. Analytical solutions for the single-particle wavefunctions are available in this case~\cite{Knap2012}; see the Supplemental Material for details of numerical calculations as well as an analytical treatment of the weak-coupling limit~\cite{SM}. The physical scales of the gas are determined by the Fermi wavevector $k_F = (6\pi^2 \bar{n})^{1/3}$, energy $E_F = \hbar^2 k_F^2/2m$, time $\tau_F = \hbar/E_F$ and temperature $T_F = E_F/k_B$, while the dimensionless parameter $k_Fa$ quantifies the impurity-gas coupling.  The time evolution of the magnitude of the decoherence function for this system is shown in Fig.~\ref{fig:vee} for various coupling strengths and temperatures. We also plot the corresponding finite-temperature absorption spectra, which are related to $v(t)$ by a Fourier transform
\begin{equation}
    \label{absorption_spectrum}
    A(\omega) = \pi^{-1}\Re \int_0^\infty \dd t\, \ee^{-\ii\omega t} v(t).
\end{equation}
Note that $A(\omega)$ is equivalent to the probability distribution of work performed by suddenly switching on the impurity potential $V_{\rm imp}(\rr)$~\cite{Silva2008,Sindona2014, FogartyPRA}. Since the properties of $v(t)$ and $A(\omega)$ have been extensively discussed in the literature~\cite{Anderson1967,Nozieres1969,Goold2011,Knap2012,Schmidt2018,Liu2020a,Liu2020b}, here we simply summarise the notable features.

Scattering from the impurity generates particle-hole excitations in the gas. For weak coupling and low temperature, these excitations are initially limited to the vicinity of the Fermi surface due to Pauli blocking (see Fig.~\ref{fig:schematic}), but repeated scattering events eventually reorganise the entire Fermi sea: this is the essence of the OC~\cite{Anderson1967}. Fig.~\ref{fig:vee} shows that at relatively short times, $\tau_F \! < \! t \! \ll \! \hbar\beta$, the OC manifests itself in a universal decoherence function $v(t)\!\sim\! \ee^{\ii w t} t^{-(\delta_F/\pi)^2}$, where $\delta_F \!=\! -\arctan(k_F a)$ is the scattering phase at the Fermi surface and $\hbar w$ is a collisional shift of the impurity's energy levels~\cite{Knap2012}. This short-time behaviour is essentially dictated by the high-frequency tails of $A(\omega)$, which describe collective excitations of the whole Fermi sea and thus are largely insensitive to temperature. At later times, $v(t)$ departs from the zero-temperature behaviour, decaying exponentially with a temperature-dependent rate for $t\gtrsim\hbar\beta$. This long-time behaviour is determined by low-energy excitations close to the Fermi surface whose distribution is highly temperature-dependent. This is seen in the dominant feature of the absorption spectra near $\omega \!=\! \Delta E$ where the zero-temperature edge singularity~\cite{Nozieres1969}, resulting from the discontinuous Fermi surface, is softened at finite temperature into a broad peak (see Fig.~\ref{fig:vee} inset). The width of the peak is determined by both the temperature and the scattering length: larger values of $|k_Fa|$ lead to a broader peak and thus a faster onset of exponential decay in the time domain. 

\begin{figure}
    \centering
     \includegraphics[width=\linewidth, trim = 0cm 10.5cm 0cm 10cm, clip]{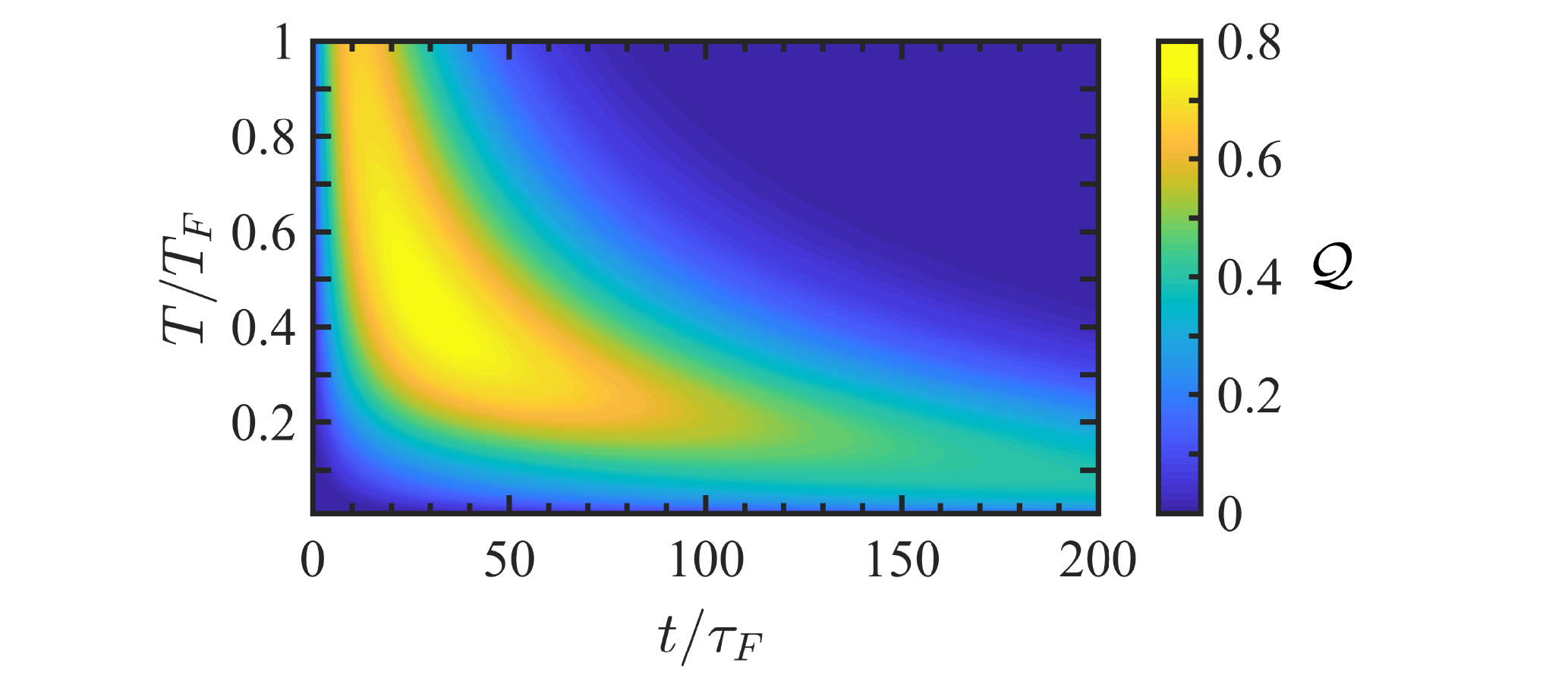}
    \caption{QSNR as a function of temperature and evolution time for $k_Fa = -0.5$. \label{fig:QSNR}}
\end{figure}

\textit{Thermometric performance.---}We now turn to the metrological implications of these features. Fig.~\ref{fig:QSNR} shows the QSNR as a function of time and temperature for $k_Fa\!=\!-0.5$. At a given temperature, the optimal measurement time corresponds to the maximum sensitivity, i.e. $\mathcal{Q}_{\rm max}\! \!=\! \!\max_{t} \mathcal{Q}(t)\! \!=\! \!\mathcal{Q}(t_{\rm max})$, which shifts to progressively later times as the temperature decreases. We find that the maximum QSNR, shown by the large yellow region in Fig.~\ref{fig:QSNR}, coincides with the relevant temperature range for current experiments~\cite{Cetina2016,Nascimbene2009,Nascimbene2010}, i.e.\ $T\!\gtrsim\! 0.1 T_F$, and good precision is retained down to the deeply degenerate regime. For example, with a coupling strength of $k_Fa\!=\!-0.5$ and a temperature of $T\!=\!0.1 T_F$ we find $\mathcal{Q}_{\rm max}\! \approx\! 0.45$, meaning that an error of $\Delta T/T \!=\! 10\%$ can be achieved with $N\!\approx\! 500$ measurements after a time  $t_{\rm max}\!\approx \!150\tau_F$, which is on the order of milliseconds for typical experimental parameters. This is eminently feasible, since a single gas sample may include thousands of independent impurities~\cite{Cetina2016} and have a lifetime of several seconds~\cite{Spiegelhalder2009,Mukherjee2017}.

\begin{figure}
    \centering
    \includegraphics[width=\linewidth, trim = 0cm 6cm 0cm 5.5cm, clip]{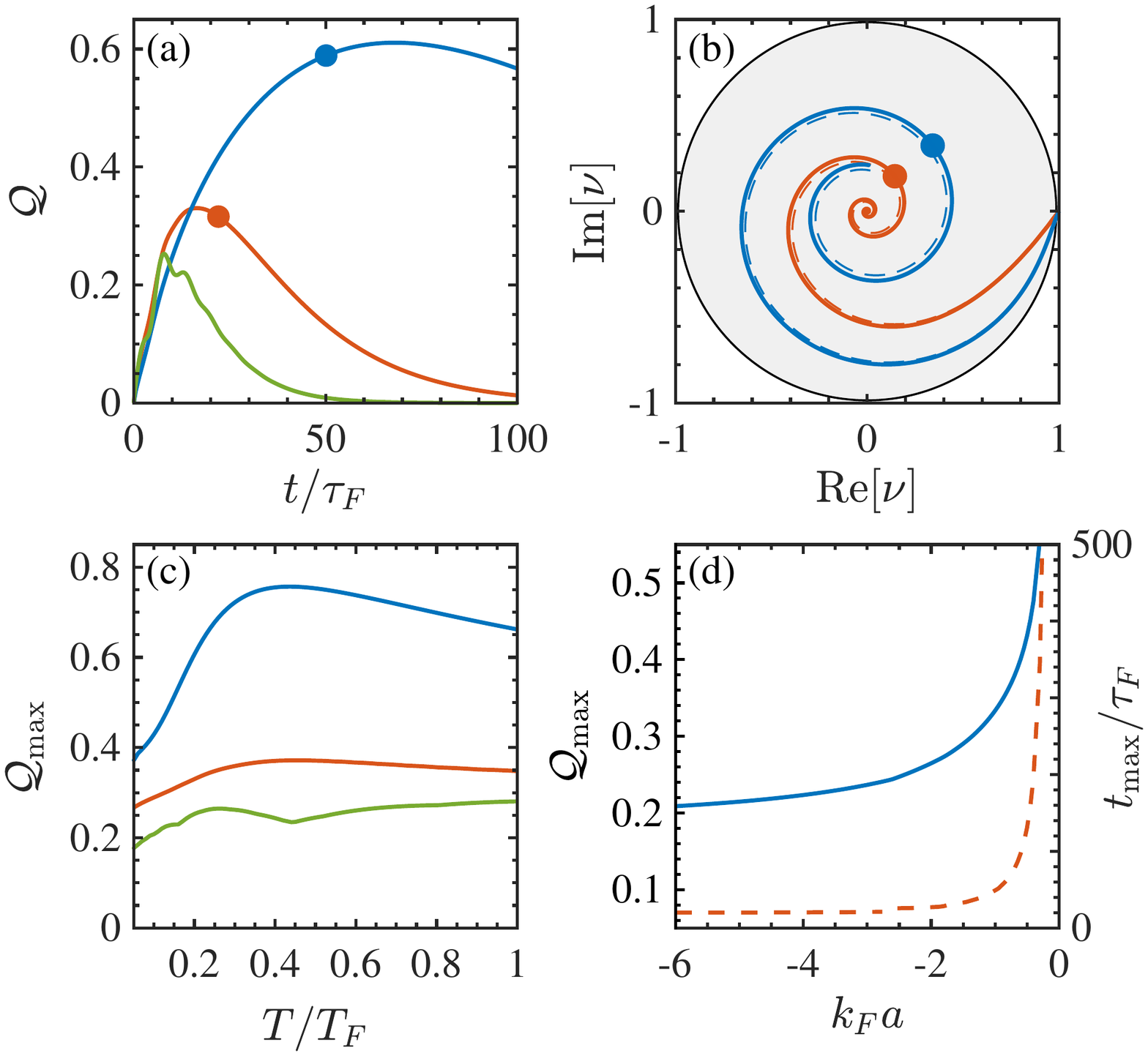}\\  
    \caption{(a) QSNR at $T\!=\!0.2T_F$ as a function of time for $k_F a \!=\! -0.5$ (blue), $k_Fa\! =\! -1.5$ (red) and $k_Fa\! =\! -6$ (green). (b) Decoherence function on the equator of the Bloch sphere for $T\!=\!0.2 T_F$ (solid lines) and $T\!=\!0.22 T_F$ (dashed lines) with $k_Fa\! =\! -0.5$ (blue) and $k_Fa\! =\! -1.5$ (red). Solid circles highlight the same instants in time in both panels. (c) Maximum sensitivity, $\mathcal{Q}_{\rm max}$, as a function of temperature for $k_F a\! =\! -0.5$ (blue), $k_Fa\! =\! -1.5$ (red) and $k_Fa \!=\! -6$ (green). (d) $\mathcal{Q}_{\rm max}$ (solid line) and corresponding measurement time (dashed line) as a function of coupling strength for $T\!=\!0.1T_F$. See text for discussion.
    \label{fig:QSNR_coupling_grid} }
\end{figure}

Naturally, the maximum precision depends on the coupling strength. In Fig.~\ref{fig:QSNR_coupling_grid}(a) we show the dynamical QSNR for various scattering lengths finding, remarkably, that weaker coupling enhances thermometric performance. This can be understood by virtue of Eq.~\eqref{QFI_qubit}, which shows that probe states with high purity, i.e.~large $|v|$, have a larger QFI. Since a state with high purity may have a sharply peaked distribution of measurement outcomes, a small parameter change is statistically easier to distinguish. Weak coupling is then preferable in light of the slower initial power-law decoherence --- due ultimately to Pauli exclusion reducing the available phase space for scattering --- which maintains purer, and therefore more sensitive, probe states. This is also illustrated in Fig.~\ref{fig:QSNR_coupling_grid}(b), which shows the path traced by the Bloch vector for two nearby temperatures and two coupling strengths. Clearly, weaker coupling ensures that the probe maintains larger purities and consequently is more sensitive to small temperature changes. In Fig.~\ref{fig:QSNR_coupling_grid}(c) we show that $\mathcal{Q}_{\rm max}$ is always larger for smaller scattering strengths, indicating that this qualitative picture holds for all temperatures.

However, this improved precision comes at the cost of measurement time. Indeed, from Fig.~\ref{fig:vee} we know that the onset of thermal behaviour is delayed by weak coupling. We quantitatively examine the thermometric implications of this in Fig.~\ref{fig:QSNR_coupling_grid}(d) where we find that both $\mathcal{Q}_{\rm max}$ and $t_{\rm max}$ diverge as $|k_F a| \! \to\! 0$. In this limit,  the universal exponent determining the decoherence rate is $(\delta_F/\pi)^2 = \mathcal{O}((k_Fa)^2)$, whereas the phase evolves as $\phi = wt$, with $w= \mathcal{O}(k_Fa)$~\cite{SM}. The QFI [Eq.~\eqref{QFI_qubit}] is thus dominated by $\mathcal{F}_T^\perp$ while the SLD [Eq.~\eqref{SLD}] is approximately $\hat{\Lambda}_T \!\approx\! \sg_\perp$, corresponding to a phase estimation protocol~\cite{Paris2009, Toth2014}. Since $w$ is temperature-dependent, small temperature variations develop over time into large, distinguishable phase differences, resulting in the asymptotic scalings $t_{\rm max}\sim |k_Fa|^{-2}$ and $\mathcal{Q}_{\rm max}\sim ~ |k_Fa|^{-1}$ at weak coupling~\cite{SM}. The universal OC physics is therefore crucial because slow, algebraic decoherence allows a long time for phase accumulation without sacrificing the purity of the probe state.

\textit{Discussion.---}Homogeneous ultracold gases represent a challenge for in situ thermometry, necessitating destructive time-of-flight measurements~\cite{Gaunt2013,Mukherjee2017}. In fermionic systems this problem is exacerbated because the Pauli exclusion principle restricts thermal excitations to a small energy window near the Fermi surface, meaning that density measurements of any kind provide little information on temperature. In contrast, our proposal to infer temperature from decoherence is designed to exploit this structure of the Fermi sea. Specifically, exclusion effects slow the decay of the impurity decoherence function, allowing for enhanced sensitivity. Moreover, our scheme is inherently non-equilibrium, thus alleviating the need for thermalisation of the probe before accurate temperature estimation is feasible. 

The sensitivity of our probe can be controlled by using a Feshbach resonance to change the scattering length. Remarkably, we have shown that the highest QSNR is obtained for \textit{weak} coupling, in contrast with the sensitivity enhancement found for thermalising probes at strong coupling~\cite{Correa2017}. Practically speaking, weak coupling reduces the number of measurements needed to achieve a given precision, albeit at the cost of increasing the measurement time (and vice versa). This tunability allows the protocol to be optimised depending on the experimental constraints at hand. It is worth noting that the impurity decoherence function exhibits a universal dependence on a small number of parameters, $k_Fa$, $E_F$ and $T$, which can each be determined via a similar interferometric protocol. For example, either $k_Fa$ or $E_F$ can be determined from the temperature-independent behaviour of $v(t)$ at short times. This may assist calibration of the thermometer and obviates the need to incorporate independent measurements of the density or scattering length --- with their associated experimental uncertainties --- into the parameter estimation procedure.

Our analysis focussed on homogeneous gases where conventional in situ thermometry is difficult. However, the same approach could in principle be applied to trapped gases with arbitrary geometry. In the Supplemental Material~\cite{SM}, we consider thermometry of the one-dimensional (1D) Fermi gas, finding similar sensitivities to the 3D case. Interestingly, the norm of the decoherence function is similar for homogeneous and harmonically trapped 1D gases (for $\omega_0t\ll \pi$, with $\omega_0$ the trap frequency~\cite{Sindona2013}). However, the complex phase of $v(t)$ is significantly modified by the presence of the harmonic trap. Since this phase is sensitive to temperature, an optimally precise temperature estimator for a harmonically confined gas should account for the trap configuration. We emphasise that our theory based on Eq.~\eqref{functional_determinant} is computationally efficient for any size and geometry, requiring only the single-particle wavefunctions.

In summary, we have proposed a minimally destructive and local thermometry protocol based on the decoherence of immersed impurities, which offers a solution to the challenge of in situ thermometry for homogeneous Fermi gases. This complements recently developed techniques based on two-photon spectroscopy~\cite{Gotlibovych2014,Shkedrov2020}. Future work could address the effect of impurity motion~\cite{Parish2016,Liu2019} and correlations between probes generated via their mutual interaction with the gas~\cite{Hangleiter2015,Mitchison2016,Planella2020}.

\textit{Acknowledgments.---} We are grateful to Alessandro Silva for unwittingly suggesting a cumulant expansion approach, and we thank Luis Correa, Michael Knap, Matteo Paris, Meera Parish, Patrick Potts, Yuval Sagi, and Artur Widera for useful comments on the manuscript. We acknowledge support from the European Research Council Starting Grant ODYSSEY (G. A. 758403), the JSPS KAKENHI-18K13507, the SFI-Royal Society University Research Fellowship scheme, the Science Foundation Ireland Starting Investigator Research Grant ``SpeedDemon" (No. 18/SIRG/5508), and the Okinawa Institute of Science and Technology Graduate University.

\bibliographystyle{apsrev4-1}
\bibliography{bibliography}

\appendix
\input{sm}

\end{document}

%% file: packages.tex
\usepackage{amsmath}
\usepackage{graphicx}
\usepackage{amsfonts}
\usepackage[dvipsnames]{xcolor}
\usepackage{upgreek}
\usepackage{amsmath}
\usepackage{amssymb}
\usepackage{graphicx} 
\usepackage{pgf}
\usepackage{epstopdf}
\usepackage{braket}
\usepackage{mathtools}
\usepackage{txfonts}
\usepackage{bbm}
\usepackage{bm}

\usepackage[colorlinks,linkcolor=blue,urlcolor=blue,citecolor=blue]{hyperref}

%% file: definitions.tex
\definecolor{islamicgreen}{rgb}{0.0, 0.56, 0.0}

\renewcommand\Re{\textrm{Re}}
\renewcommand\Im{\textrm{Im}}
\newcommand\Tr{\textrm{Tr}}

\renewcommand\ket[1]{\left|\textstyle{#1}\right\rangle}
\renewcommand\bra[1]{\left\langle\textstyle{#1}\right|}

\newcommand{\ii}{\ensuremath{\mathrm{i}}}
\newcommand{\ee}{\ensuremath{\mathrm{e}}}
\newcommand{\dd}{\mathrm{d}}

\newcommand{\rr}{{\bf r}}

\newcommand{\sg}{\ensuremath{\hat{\sigma}}}

\renewcommand{\mathcal}[1]{\text{\usefont{OMS}{cmsy}{m}{n}#1}}


%% file: sm.tex
\clearpage
\section*{Supplemental Material}

\setcounter{secnumdepth}{2}
\setcounter{equation}{0}
\setcounter{figure}{0}
\renewcommand{\theequation}{S\arabic{equation}}
\renewcommand{\thefigure}{S\arabic{figure}}

\subsection{Calculation of the decoherence function}
\label{app:calculation_details_3D}

In this section, we give details on our calculation of the decoherence functions displayed in the main text based on the Levitov formula [Eq.~\eqref{functional_determinant} main text]
\begin{equation}
    \label{Levitov_SM}
        v(t) = \det \left[1 - \hat{n} + \hat{n}\ee^{\ii\hat{h}_0t/\hbar}\ee^{-\ii\hat{h}_1t/\hbar}\right ].
\end{equation}
We work in spherical coordinates with the impurity placed at $r=0$ and impose hard-wall boundary conditions at a radius $r=R$. Assuming low temperatures and $s$-wave scattering, only atoms with zero angular momentum are perturbed by the impurity. Higher partial-wave states thus do not contribute to the determinant. 

The $s$-wave eigenfunctions of the perturbed and unperturbed Hamiltonians are denoted by $\psi_n(\rr)$ and $\psi_n'(\rr)$ respectively, such that $\hat{h}_0\psi_n = E_n \psi_n$ and $\hat{h}_1 \psi'_n = E_n'\psi_n'$.  They are given explicitly by~\cite{Knap2012}
\begin{align}
    \label{3D_solutions}
    \psi_n(\rr) & = \sqrt{\frac{1}{2\pi R}} \frac{\sin(k_{n}r)}{ r},\\
    \label{3D_perturbed_solutions}
    \psi'_n(\rr) & = A_n\sqrt{\frac{1}{2\pi R}} \frac{\sin(k_n' r + \delta_n)}{ r},
\end{align}
where $k_nR = k_n'R + \delta_n = n\pi$ for $n=1,2,\ldots$, while ${E_n = \hbar^2 k_n^2/2m}$ and $E_n'=\hbar^2k_n'^2/2m$. The scattering phase is determined by the equation
\begin{equation}
    \label{scattering_phase_3D}
    \tan(\delta_n) = -k_n'a,
\end{equation}
with $A_n = \left[1+\sin(2\delta_n)/2k_n'R\right]^{-1/2}$. Here we have assumed negative scattering lengths so that no bound state arises.

The Fermi energy is determined by $E_F = \hbar^2 k_F^2/2m = \pi^2\hbar^2 N_s^2/2mR^2$, where $N_s$ is the number of atoms in $s$ states at $T=0$. We work at fixed 3D density and thus hold $E_F$ fixed, which is equivalent to fixing the ratio $N_s/R = \sqrt{2mE_F}/\pi\hbar$. All of our results are scaled to the thermodynamic limit by increasing $N_s$ until convergence is achieved. On the timescales we consider, $N_s$ on the order of a few hundred is sufficient. 

The determinant in Eq.~\eqref{Levitov_SM} is infinite-dimensional in principle, but at any given density and temperature it can be computed  to high accuracy within a finite basis set. The size $N$ of the unperturbed basis set $\{\psi_n\}_{n=1}^N$ is fixed by the temperature and $N_s$. For each value of $T$, we find the chemical potential by solving $\Tr[\hat{n}] = N_s$ with a very large basis ($\sim~10^4$~states). We then truncate to $N$ states such that $|\Tr[\hat{n}] - N_s|<\epsilon$, where $\epsilon$ is a small tolerance. The size of the perturbed basis set $\{\psi'_n\}_{n=1}^{N'}$ is then fixed by the requirement of unitarity: $\sum_{n=1}^{N'}|\langle \psi_m|\psi'_{n}\rangle|^2 > 1-\epsilon$ for all $m\leq N$. We find that $\epsilon = 10^{-4}$ is sufficient to obtain good convergence. 

The Fisher information is given by [Eq.~\eqref{QFI_qubit} main text]
\begin{equation}
    \label{QFI_SM}
    \mathcal{F}_T^Q  = \frac{1}{1-|v|^2}\left(\frac{\partial |v|}{\partial T}\right)^2 + |v|^2 \left(\frac{\partial \phi}{\partial T}\right)^2.
\end{equation}
This is evaluated via a finite-difference approximation to the temperature derivatives with numerical increments of $\delta T/T = 10^{-2}$.

\subsection{Cumulant expansion at weak coupling}
\label{app:cumulant_expansion}

In this section, we detail an analytical calculation of the decoherence function valid for weak coupling, $k_Fa\ll 1$, and low temperatures, $T\ll T_F$. We start from the many-body representation of the decoherence function, $v=|v|\ee^{\ii\phi}$, given by [Eq.~\eqref{decoherence_function} main text]
\begin{equation}
\label{decoherence_func_manybody}
   v(t) = \Tr_E \left[\ee^{\ii\hat{H}_{1}t/\hbar}\ee^{-\ii\hat{H}_{0}t/\hbar}\hat{\rho}_{E}(T)\right].
\end{equation}
For the homogeneous 3D gas considered in the main text, we may take $\hat{H}_1 = \hat{H}_0 + \hat{V}$, where 
\begin{align}
\label{H_0_many_body}
    \hat{H}_0 & = \sum_n E_n \hat{c}_n^\dagger \hat{c}_n,\\
\label{V_many_body}
    \hat{V} & = \sum_{l,n} V_{ln} \hat{c}_l^\dagger \hat{c}_n.
\end{align}
Here, $\hat{c}_n^\dagger$ creates a fermion in the $s$-wave state $\psi_n(\rr)$ given by Eq.~\eqref{3D_solutions}, $E_n = \hbar^2 k_n^2/2m$ is the corresponding energy and $V_{ln}$ is the interaction matrix element
\begin{equation}
    \label{interaction_matrix_element}
    V_{ln} = \int \dd^3\rr \, \psi^*_l(\rr) V_{\rm imp}(\rr) \psi_n(\rr) = \frac{\hbar^2 a}{m R} k_l k_n,
\end{equation}
where we have invoked the infinite-mass approximation for the impurity. All higher partial-wave states are unaffected by $\hat{V}$ and therefore do not contribute to the decoherence function.

We proceed by expressing the complex conjugate of Eq.~\eqref{decoherence_func_manybody} as a time-ordered exponential, which can then be expanded in terms of time-ordered cumulants~\cite{Kubo1962} as
\begin{align}
    \label{cumulant_expansion}
    |v|\ee^{-\ii\phi} & = \left \langle \overset{\leftarrow}{\rm T} \exp \left [ \int_0^t\dd t' \, \frac{\hat{V}(t')}{\ii \hbar}  \right] \right\rangle \\ 
    & \approx \exp \left [  \left\langle  \int_0^t\dd t' \,  \frac{\hat{V}(t')}{\ii \hbar} \right\rangle_{\rm c}  + \frac{1}{2}\overset{\leftarrow}{\rm T} \left\langle \left( \int_0^t\dd t' \, \frac{\hat{V}(t')}{\ii \hbar}\right)^2\right\rangle_{\rm c} \right].\notag
\end{align}
Above, $\hat{V}(t) = \ee^{\ii \hat{H}_0 t/\hbar} \hat{V} \ee^{-\ii \hat{H}_0 t/\hbar}$ is the perturbation in the interaction picture, $\overset{\leftarrow}{\rm T}$ indicates chronological time ordering, $\langle \bullet\rangle$ denotes an average with respect to the initial thermal state $\hat{\rho}_E(T)$ while $\langle \bullet\rangle_{\rm c}$ denotes the corresponding cumulant. On the second line of Eq.~\eqref{cumulant_expansion}, we have neglected terms of order $\mathcal{O}(\hat{V}^3)$ in the exponent. 

The first cumulant is found to be
\begin{align}
    \label{first_term_exponent}
\left\langle  \int_0^t\dd t' \,  \frac{\hat{V}(t')}{\ii \hbar} \right\rangle_{\rm c} = \frac{t}{\ii \hbar}\left \langle \hat{V}\right \rangle = -\ii w t ,
\end{align}
where we recognise $\hbar w = \langle \hat{V}\rangle$ as the first-order energy shift (the mean work done) due to the perturbation. Using the thermal average $\langle \hat{c}_l^\dagger \hat{c}_n\rangle = f(E_n)\delta_{ln}$, we obtain the shift explicitly as
\begin{equation}
    \label{first_order_shift}
    \hbar w = \frac{2a}{R} \sum_n E_n f(E_n) = 2a \int_0^\infty\dd E\, D_s(E) E f(E),
\end{equation}
where $D_s(E) = R^{-1}\sum_n\delta(E-E_n)$ is the $s$-wave density of states per unit radius, which for $R\to \infty$ becomes
\begin{equation}
\label{density_of_states}
    D_s(E) = \frac{1}{\pi\hbar} \sqrt{\frac{m}{2E}}.
\end{equation}
It is notable that Eq.~\eqref{first_order_shift} is proportional to the average energy density of $s$-wave fermions in the gas. We therefore see already how the temperature is imprinted onto the phase of the decoherence function.

In order to model the loss of phase coherence, we must consider the second cumulant in Eq.~\eqref{cumulant_expansion}. Using the standard relation for fermionic Gaussian states~\cite{Cahill1999}, $\langle c^\dagger_j \hat{c}_k \hat{c}^\dagger_l \hat{c}_n\rangle_{\rm c} \equiv \langle c^\dagger_j \hat{c}_k \hat{c}^\dagger_l \hat{c}_n\rangle - \langle c^\dagger_j \hat{c}_k \rangle \langle \hat{c}^\dagger_l \hat{c}_n\rangle = f(E_n)[1-f(E_l)]\delta_{jn}\delta_{kl}$, we obtain
\begin{align}
\label{second_cumulant}
   & \frac{1}{2}\overset{\leftarrow}{\rm T} \left\langle \left( \int_0^t\dd t' \, \frac{\hat{V}(t')}{\ii \hbar}\right)^2\right\rangle_{\rm c}  \\ & 
    = \int_0^t\dd t'\int_0^{t'}\dd t''\sum_{l,n} \left(\frac{V_{ln}}{\ii\hbar}\right)^2 \ee^{-\ii (E_n-E_l)t''/\hbar} f(E_l)[1-f(E_n)].\notag
\end{align}
Diagonal terms with $l=n$ lead to a contribution proportional to $t^2$ which vanishes as $R^{-1}$ and is therefore negligible in the large-system limit, although this term remains relevant for trapped gases~\cite{Sindona2013}. The remaining elements with $l\neq n$ can be grouped according to their real and imaginary parts, corresponding to the dephasing function $\Gamma(t)$ and the second-order phase shift $\Phi(t)$, respectively. These are written compactly as
\begin{align}
    \label{dephasing_function}
    \Gamma(t) & = \fint_{-\infty}^\infty \dd \omega\, \frac{J(\omega)}{\omega^2}\left[1-\cos(\omega t)\right],\\
\label{second_order_shift}
    \Phi(t)& = \fint_{-\infty}^\infty \dd \omega\, \frac{J(\omega)}{\omega^2}[\omega t - \sin(\omega t)],
    \end{align}
where $\fint$ denotes a principal-value integral excluding $\omega=0$, and we defined the spectral density \begin{align}
    \label{spectral_density}
    J(\omega) & = \frac{1}{\hbar}\sum_{l,n} V_{ln}^2 f(E_l) [1-f(E_n)]\delta(\hbar\omega + E_l - E_n),
\end{align}
which represents the coupling strength to particle-hole excitations of energy $\hbar\omega$, weighted by their finite-temperature density of states.

To make further progress, we take the continuum limit using Eq.~\eqref{density_of_states} to obtain
\begin{equation}
    \label{spectral_density_integral}
    J(\omega) = \frac{\alpha}{\hbar E_F} \int_0^\infty\dd E \sqrt{E(E+\hbar\omega)} f(E)[1-f(E+\hbar\omega)],
\end{equation}
where $\alpha = (k_Fa/\pi)^2$ is a dimensionless coupling strength. Considering low temperatures, such that ${T\ll T_F}$ and $\mu\approx E_F$, it is clear that $J(\omega)$ is exponentially suppressed for $\hbar\omega \lesssim -E_F$, while $J(\omega) \sim \alpha \sqrt{\omega\tau_F}$ for $\hbar\omega \gtrsim E_F$. Comparing with Eq.~\eqref{dephasing_function}, we see that this high-frequency part of $J(\omega)$ thus leads to a small contribution to $\Gamma(t)$ on the order of $\alpha$, which does not grow in time for $t\gg \tau_F$. The long-time dephasing dynamics is instead dominated by the behaviour of $J(\omega)$ at low frequencies, $\hbar |\omega|\ll E_F$. In this regime of low frequencies and temperatures, the function $f(E)[1-f(E+\hbar\omega)]$ is sharply peaked near $E=E_F$. We may therefore make the replacement $\sqrt{E(E+\hbar\omega)} \to \sqrt{E_F(E_F+\hbar\omega)} \approx E_F$ and send the lower integration limit to $-\infty$ with negligible error. The integral may then be carried out, yielding the approximation
\begin{equation}
    \label{spectral_function_Ohmic}
    J(\omega) \approx \tfrac{1}{2} \alpha \omega \left[1 + \coth(\beta\hbar \omega/2)\right],
\end{equation}
which is valid for frequencies $|\omega|< \Lambda$, where $\Lambda$ is an ultraviolet (UV) cutoff on the order of the Fermi energy, i.e.~$\hbar\Lambda\lesssim E_F$.  Eq.~\eqref{spectral_function_Ohmic} takes the form of an Ohmic spectral density, with the appearance of a bosonic occupation factor signalling that low-frequency particle-hole excitations behave like an effective bosonic bath for the impurity. However, an important distinction from the standard bosonic environment typically considered, e.g.~in the independent boson model~\cite{Bruderer_2006,Razavian2019}, is that the average perturbation $\langle \hat{V}\rangle$ does not vanish and indeed carries valuable information on temperature that can be extracted through interferometry.

Returning now to Eqs.~\eqref{dephasing_function} and~\eqref{second_order_shift}, and using the approximation in Eq.~\eqref{spectral_function_Ohmic}, we obtain
\begin{align}
\label{re_Gamma}
\Gamma(t) & = \frac{\alpha}{2\ii} \int_0^t\dd t' \fint_{-\Lambda}^\Lambda \dd \omega\,\coth(\beta\hbar\omega/2)\ee^{\ii\omega t'},\\
\label{im_Gamma}
\Phi(t) & = \frac{\alpha}{2} \fint_{-\Lambda}^\Lambda \dd\omega\,\frac{\omega t - \sin(\omega t)}{\omega} ,
\end{align}
where we have neglected frequencies beyond the UV cutoff $\Lambda\sim E_F/\hbar$ since, as discussed above, these make a negligible contribution to $\Gamma(t)$ for times $t\gg \tau_F$. For $\Lambda t \gg 1$, we find $\Phi(t) \approx \alpha (\Lambda t + \pi/2)$, which describes a small additional phase shift of order $\Phi \sim \sqrt{\alpha}w t$ [note that $w = \mathcal{O}(\sqrt{\alpha})$], which will be neglected henceforth. We note that the behaviour of $J(\omega)$ for high frequencies, $\omega>\Lambda$, which we have ignored above, leads to a UV divergence in Eq.~\eqref{second_order_shift}. This underlines the fundamentally non-perturbative nature of the Anderson orthogonality catastrophe: all orders in perturbation theory are needed to achieve a better approximation than just the first-order shift given by Eq.~\eqref{first_order_shift}, with the correct result at $T=0$ given by Fumi's theorem (see Ref.~\cite{Schmidt2018} for details).

It remains for us to find the dephasing function. Eq.~\eqref{re_Gamma} may be evaluated approximately as follows:
\begin{align}
     \label{coth_Fourier}
\Gamma(t)     &\approx \frac{\alpha}{2\ii}\int_0^t\dd t' \left[ \fint_{-\infty}^\infty\dd \omega\, \coth(\beta\hbar\omega/2) \ee^{\ii\omega t'}
    \right.  \\
    \label{Sok_Plem}
     & \qquad  \qquad \left. - \int_\Lambda^\infty \dd \omega\, \ee^{\ii\omega t'} - \int_{-\infty}^{-\Lambda} \dd \omega\, \left(-\ee^{\ii\omega t'}\right)\right ] \\
     \label{dephasing_rate_explicit}
     & =  \alpha\int_0^t\dd t'\,\left[\frac{\pi \coth(\pi t'/\hbar \beta)}{\hbar\beta} - \frac{\cos(\Lambda t')}{t'}\right]\\
     \label{dephasing_function_explicit}
     & =  \alpha\left\lbrace\ln \left [ \frac{ \hbar \Lambda \beta}{ \pi}\sinh\left(\frac{\pi t}{\hbar \beta}\right) \right] - {\rm Ci}(\Lambda t) + \gamma_{\rm E}\right\rbrace.
\end{align}
On lines~\eqref{coth_Fourier} and~\eqref{Sok_Plem}, we have partitioned the integration domain as shown and made the approximation $\coth(\beta\hbar\omega/2) \approx {\rm sign}(\omega)$ for $|\omega|>\Lambda$, which is justified so long as $\beta\hbar\Lambda\gg 1$. The integral on line~\eqref{coth_Fourier} is essentially the (principal-value component of the) Fourier transform of $\coth(\beta\hbar\omega/2)$. This is computed by continuing the integrand to complex frequencies and closing the integration contour in the upper half-plane, resulting in a geometric sum over the residues of the poles at $\omega = \ii \omega_n$ for $n = 0,1,2,\ldots$, where $\omega_n = 2n\pi/\hbar\beta$ are bosonic Matsubara frequencies. The principal-value component of the integral along the real line, which excludes the origin, is found after subtracting  one half of the residue (times $\ii 2\pi$) at $\omega=0$. The remaining two integrals on line~\eqref{Sok_Plem} are easily computed with the help of the Sokhotski-Plemelj theorem. The resulting expression, Eq.~\eqref{dephasing_rate_explicit}, can then be integrated exactly over time to yield Eq.~\eqref{dephasing_function_explicit}, with $\gamma_{\rm E}$ the Euler constant and ${\rm Ci}(z) = -\int_{z}^\infty \dd x \cos(x)/x$ the cosine integral function. This cutoff-dependent term regulates the solution, which would otherwise diverge at short times, but is negligible on the timescales of interest since ${\rm Ci}(\Lambda t) \approx 0$ for $\Lambda t\gg 1$.

\begin{figure}
    \centering
    \includegraphics[width=\linewidth, trim = 0cm 8cm 0cm 8cm]{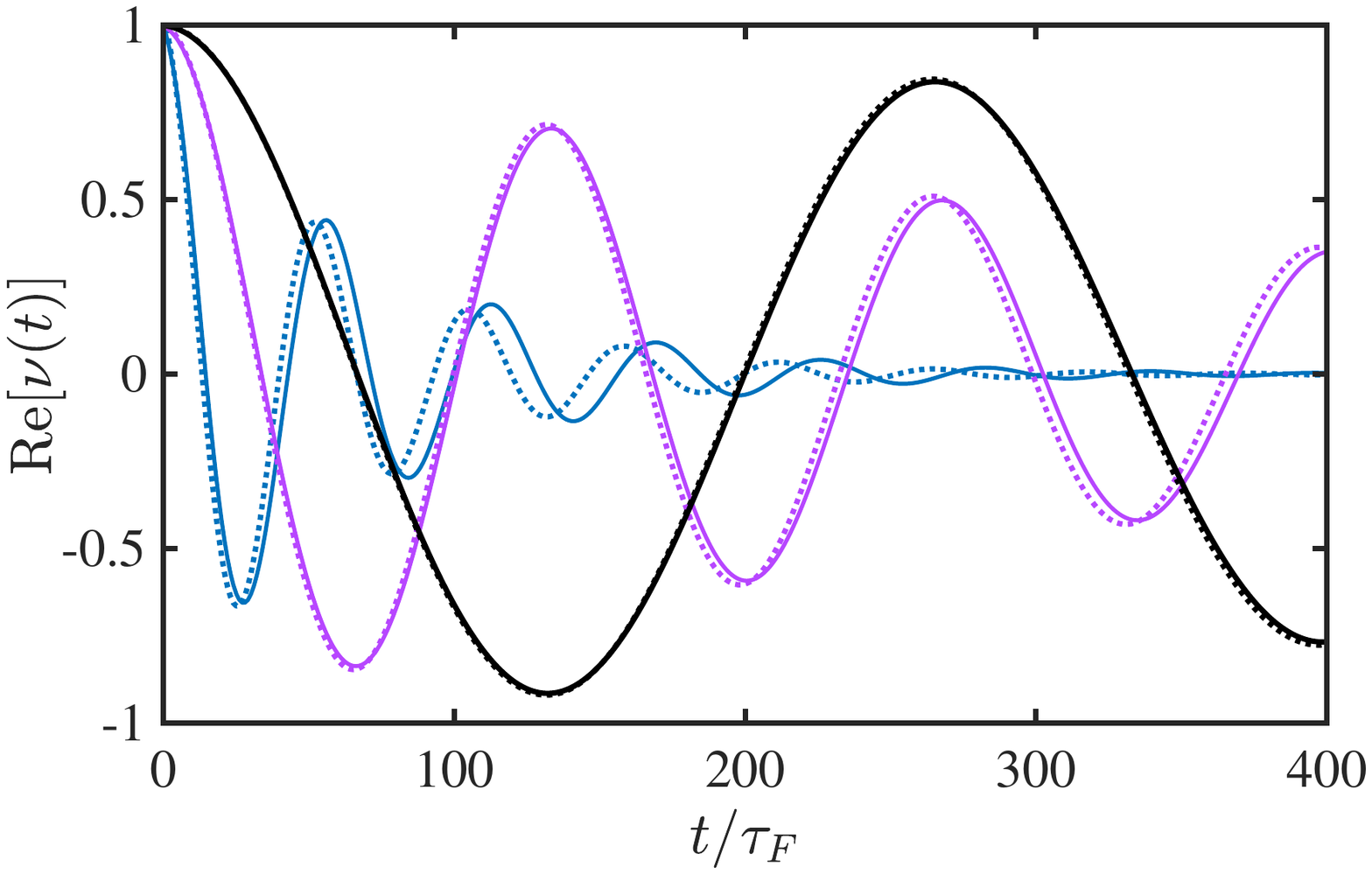}
    \caption{Comparison between the (real part of the) exact decoherence function computed from the Levitov formula in Eq.~\eqref{Levitov_SM} (solid) and the weak-coupling approximation given by Eq.~\eqref{decoherence_function_approx} (dotted), for temperature $T=0.2T_F$ and various weak coupling strengths $k_Fa = -0.5$ (blue), $k_Fa = -0.2$ (purple) and $k_Fa = -0.1$ (black).}
    \label{fig:weak_coupling_approx}
\end{figure}

Putting everything together and neglecting factors of order unity (in particular, $(\Lambda\tau_F)^{-\alpha} \approx 1$), we find the decoherence function to be well approximated by
\begin{equation}
    \label{decoherence_function_approx}
    v(t) \approx \ee^{\ii w t} \left [ \frac{ \hbar \beta}{ \pi \tau_F}\sinh\left(\frac{\pi t}{\hbar \beta}\right) \right]^{-\alpha}.
\end{equation}
This result is valid for weak coupling, $k_F a\ll 1$, low temperatures, $T\ll T_F$, and times $t \gg \tau_F$. For times less than the thermal time, $\tau_F \ll  t \ll \hbar\beta$, we find algebraic decoherence with exponent $\alpha = (k_Fa/\pi)^2$. At longer times, $t\gtrsim  \hbar \beta$, the decoherence dynamics crosses over to exponential decay, $|v|\sim \ee^{-\gamma t}$, with decay rate $\gamma = \pi\alpha/\hbar\beta$. These results agree with the analysis of Ref.~\cite{Schmidt2018} based on bosonisation techniques, while additionally describing the temperature dependence of the phase in the weak-coupling regime [Eq.~\eqref{first_order_shift}]. At stronger coupling strengths, the exponent generalises to $\alpha = (\delta_F/\pi)^2$, with $\delta_F = -\arctan(k_Fa)$ the scattering phase at the Fermi surface~\cite{Schmidt2018}.

\begin{figure}
    \centering
    \includegraphics[width=\linewidth, trim = 0cm 8cm 0cm 8cm]{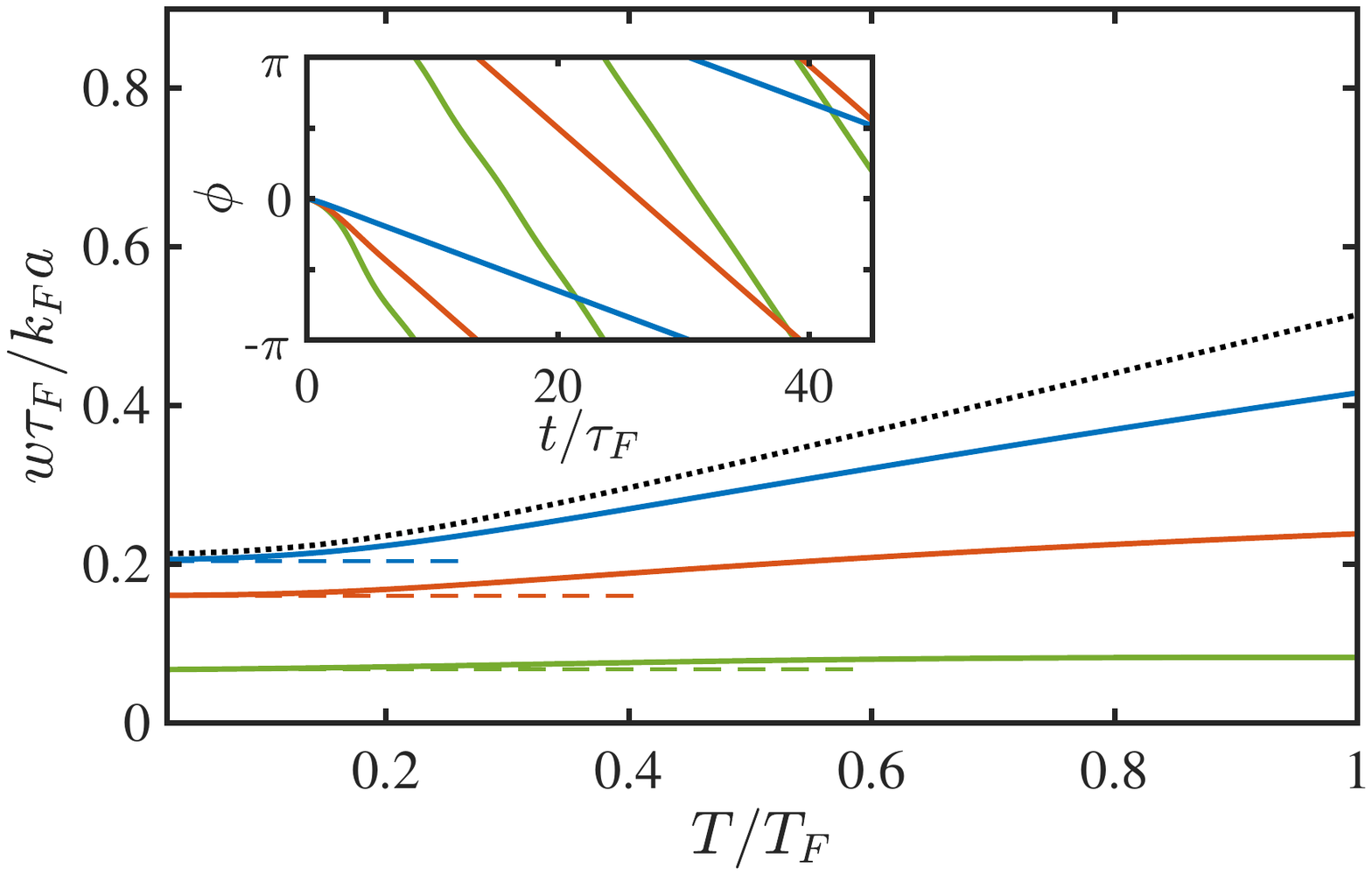}
    \caption{Phase of the decoherence function $v(t) = |v|\ee^{\ii\phi}$ as a function of time, for coupling strengths ${k_Fa=-0.5}$ (blue), $k_Fa = -1.5$ (red), and $k_Fa = -6$ (green). The inset shows the phase at fixed temperature $T = 0.1T_F$ as a function of time. The main panel shows the time derivative of the phase, $w = \dd \phi/\dd t$, at a fixed time $t=40\tau_F$ as a function of temperature (solid lines). The corresponding $T=0$ values given by Fumi's theorem are shown by the dashed lines. The dotted black line shows the approximation given by Eq.~\eqref{first_order_shift}.}
    \label{fig:phases}
\end{figure}

In Fig.~\ref{fig:weak_coupling_approx} we compare this approximate result
to the exact solution given by Eq.~\eqref{Levitov_SM}. We find that Eq.~\eqref{decoherence_function_approx} is an excellent approximation for couplings on the order of $k_F a = 0.1$ or less, even at temperature $T=0.2T_F$. As the coupling is increased, a quantitative discrepancy with the exact solution emerges already for $k_F a=0.5$. However, the qualitative features predicted by Eq.~\eqref{decoherence_function_approx} can be observed over a wide range of coupling strengths. In particular, a key prediction of our approach is that the phase of the decoherence function evolves linearly in time, i.e. $\phi(t) = wt $ with $w = \dd \phi/\dd t$ a constant. As can be seen in the inset of Fig.~\ref{fig:phases}, the phase indeed grows linearly apart from an initial transient that appears at stronger coupling (green line in Fig.~\ref{fig:phases} inset). In the main panel of Fig.~\ref{fig:phases} we show how the rate of phase accumulation, $w$, depends on temperature at a fixed time for several coupling strengths. We find the strongest temperature dependence at weak coupling, where the exact numerical results converge to the analytical prediction given by Eq.~\eqref{first_order_shift}. At stronger coupling, $w$ becomes almost independent of temperature, being dominated by the ground-state energy shift (dashed lines in Fig.~\ref{fig:phases}) given by Fumi's theorem~\cite{Schmidt2018} as $\hbar w \approx -\int_0^{E_F}\dd E\, \delta(E)/\pi$, with $\delta(E)$ the energy-dependent scattering phase (dashed lines in Fig.~\ref{fig:phases}). 

These results reinforce the idea that strong collisional coupling, which generates particle-hole excitations over a range of energies much greater than $k_BT$, tends to mask thermal effects. In contrast, weak coupling predominantly creates excitations close to the Fermi surface with energies $\hbar\omega \lesssim k_B T$, which are therefore sensitive to the shape of the initial thermal distribution. 

\subsection{Thermometric precision at weak coupling}
\label{app:weak_coupling_precision}

In this section, we analyse thermometric precision in the weak-coupling limit using the approximations developed in the previous section. Specifically, we use Eq.~\eqref{decoherence_function_approx} in the expression for the QFI given by Eq.~\eqref{QFI_SM}. We first note that the integral defining $w$ in Eq.~\eqref{first_order_shift} can be computed exactly, yielding $\hbar w = -k_Fa\,{\rm Li}_{3/2}(-\ee^{\beta\mu})/\sqrt{4\beta^3E_F} $, with ${\rm Li}_n(z)$ the polylogarithm of order $n$, so that $(\partial_T\phi)^2 = \mathcal{O}(\alpha)$, where $\alpha = k_Fa/\pi \ll 1$. Meanwhile, Eq.~\eqref{decoherence_function_approx} implies that $(\partial_T|v|)^2 = \mathcal{O}(\alpha^2)$, which is of higher order in the small parameter. Moreover, $|v|$ is a monotonically decreasing function of time, whereas $\phi$ is proportional to $t$. We may therefore safely neglect the contribution proportional to $(\partial_T|v|)^2$ in Eq.~\eqref{QFI_SM} [i.e., $\mathcal{F}^{\parallel}_T$  in Eq.~\eqref{QFI_qubit}].

Under this assumption, the QSNR $\mathcal{Q} = T \sqrt{\mathcal{F}^Q_T}$ is given by
\begin{equation}
    \label{Fisher_weak_coupling}
    \mathcal{Q} \approx t|v(t)| T \frac{\partial w}{\partial T}.
\end{equation}
We immediately notice that the accumulation of phase $\phi = w t$ over time yields a linear growth of $\mathcal{Q}$, which is counteracted by the loss of purity as $|v(t)|$ decays to zero. The competition between these two effects determines the optimum measurement time and corresponding sensitivity, $\mathcal{Q}_{\rm max} = \mathcal{Q}(t_{\rm max})$, which are found by maximising the QSNR, $\partial \mathcal{Q}/\partial t\rvert_{t=t_{\rm max}} = 0$. The solution of the maximisation problem is
\begin{equation}
    \label{t_max_solution}
   \frac{T_F}{\pi\alpha T}  = \frac{\pi t_{\rm max}}{\hbar \beta} \coth\left(\frac{\pi t_{\rm max}}{\hbar \beta}\right) \approx \frac{\pi t_{\rm max}}{\hbar \beta},
\end{equation}
where in the second equality we assumed $\pi\alpha T\ll T_F $. This shows that the optimum measurement time diverges as $\alpha^{-1}$ in the weak-coupling limit. Plugging the above solution back into Eq.~\eqref{Fisher_weak_coupling}, we get 
\begin{align}
    \label{Q_max_solution}
    \mathcal{Q}_{\rm max} & = \frac{\hbar}{\pi k_B \alpha} \left(\frac{T_F}{\pi T}\right)^{1-\alpha}  \left[ \sinh\left(\frac{T_F}{\pi \alpha T} \right)\right]^{-\alpha} \frac{\partial w}{\partial T}\notag \\
    & \approx \frac{
\hbar T_F \ee^{-T_F/\pi T}}{\pi^2 k_B\alpha T}  \frac{\partial w}{\partial T},
\end{align}
where in the second equality we again used ${\pi\alpha T \ll T_F}$ to replace the $\sinh$ function by an exponential, and we neglected the small residual exponent $\alpha\ll 1$. Since $w = \mathcal{O}(\sqrt{\alpha})$, it follows that the maximum sensitivity diverges more slowly, as $\alpha^{-1/2}$. We also note that the sensitivity is exponentially suppressed as $T\to 0$, forbidding finite thermometric precision as absolute zero is approached.

\subsection{One-dimensional and harmonically trapped systems}
\label{app:harmonic_trap_1D}

In this section, we discuss how reduced spatial dimensionality and the presence of a harmonic trap affect the sensitivity of our dephasing thermometer. To be concrete, we focus on a one-dimensional (1D) system. In this case, the impurity-gas interaction is described in the pseudo-potential approximation by $V_{\rm imp}(x) = \lambda \delta(x)$, with $\lambda = -\hbar^2/m_{\rm red}a$~\cite{Olshanii1998}. Note that the interaction strength is inversely proportional to the scattering length in 1D. We consider a tightly localised impurity at $x=0$ so that $m_{\rm red} = m$ and $|\chi(x)|^2\approx \delta(x)$.

Let us first consider a homogeneous gas and impose hard-wall boundary conditions at $x=\pm L/2$. Only the wavefunctions with even symmetry are perturbed by the presence of the impurity at $x=0$; the odd solutions thus do not contribute to the determinant in Eq.~\eqref{Levitov_SM}. The even eigenfunctions of $\hat{h}_0$ and $\hat{h}_1$ are respectively found to be
\begin{align}
    \label{1D_box_eigenfunctions}
    \psi_n(x) & = \sqrt{\frac{2}{L}} \cos\left(k_{n}x\right),\\
    \psi'_n(x) & = B_n\sqrt{\frac{2}{L}} \cos(k_n' x \pm \delta_n),
\end{align}
where the plus (minus) sign pertains to $x>0$ ($x<0$) and $k_nL = k_n'L + 2\delta_n = (2n-1)\pi$ for $n=1,2,\ldots$, while the corresponding energies are $E_n = \hbar^2 k_n^2/2m$ and $E_n'=\hbar^2k_n'^2/2m$. Note that here we assume negative scattering length; for $a>0$ the $n=1$ solution is a bound state that must be accounted for separately. The scattering phase is determined by the equation
\begin{equation}
\label{scattering_phase_1D}
    \tan(\delta_n) = \frac{1}{k_n'a},
\end{equation}
while $B_n = [1- \sin(2\delta_n)/k_n'L]^{-1/2}$. Similar to the 3D case, holding the density fixed leads to the relation $(2N_e-1)/L = \sqrt{2mE_F}/\pi\hbar$, where $N_e$ is the number of atoms in even states. We follow the same procedure as in the 3D case to find the truncated bases for a given value of $N_e$, and then scale to the thermodynamic limit by increasing $N_e$ until convergence is reached.

\begin{figure}
    \centering
    \includegraphics[width=\linewidth, trim = 0cm 8cm 0cm 8cm,clip]{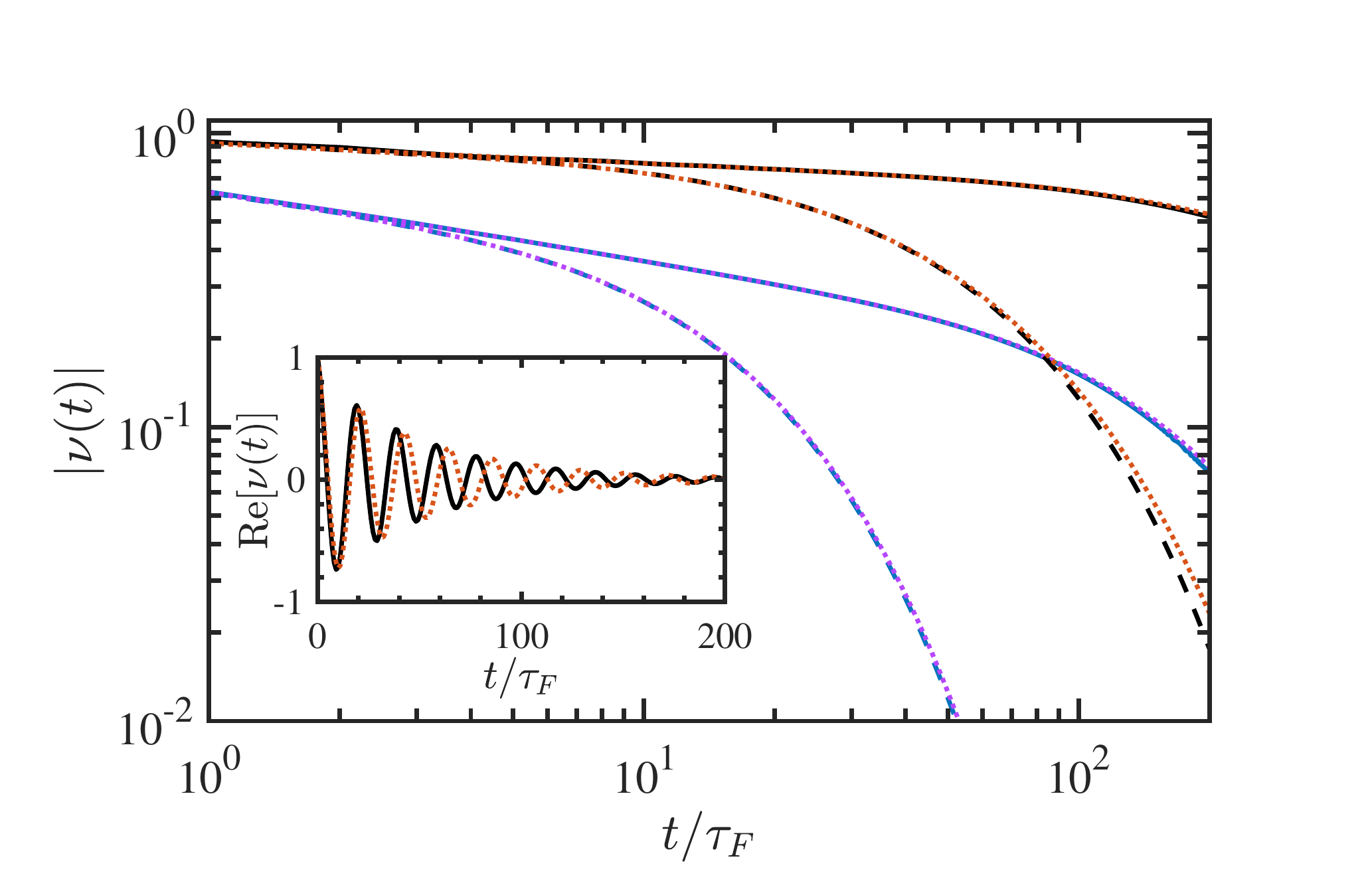}
    \caption{Comparison of decoherence functions for a homogeneous and a harmonically trapped gas in 1D. The main panel shows the absolute value for the homogeneous gas at couplings $k_Fa = -1$ (black) and $k_Fa = -0.01$ (blue) with temperatures $T = 0.01T_F$ (solid) and $T = 0.1T_F$ (dashed). Red and purple dotted lines show the corresponding results for a harmonically confined gas with $\hbar\omega_0/E_F = 2.5\times 10^{-3}$. The inset displays the real part of the decoherence function for $k_Fa=-1$ and $T=0.1T_F$.}
    \label{fig:harmonic_1D}
\end{figure}

\begin{figure}
    \centering
    \includegraphics[width=\linewidth, trim = 0cm 7.8cm 0cm 8.2cm,clip]{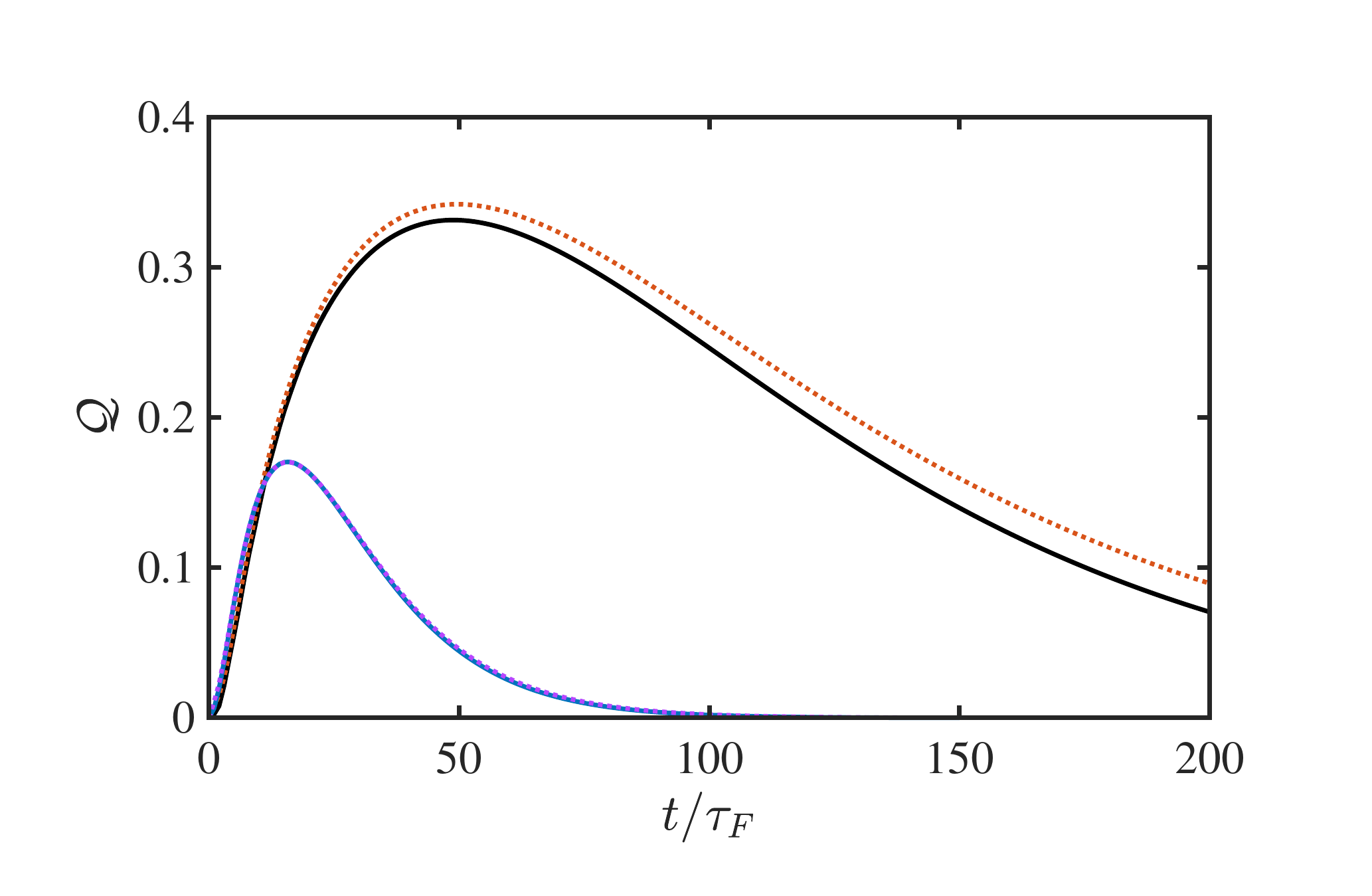}
    \caption{Comparison of the QSNR for a homogeneous (solid) and a harmonically trapped (dotted) gas in 1D at temperature $T=0.1$ and coupling strengths $k_Fa = -1$ (black, red) and $k_Fa = -0.01$ (blue, purple).}
    \label{fig:QSNR_1D}
\end{figure}

Some examples of the decoherence function are plotted in Fig.~\ref{fig:harmonic_1D} for different coupling strengths and temperatures. The qualitative behaviour is similar to the 3D case. The short-time behaviour of $v(t)$ is an oscillatory power-law decay that passes over to exponential decay after a time on the order of $\hbar/k_B T$. In Fig.~\ref{fig:QSNR_1D} we show the QSNR, finding similar results to the 3D case. In particular, we find again that weaker coupling, corresponding in 1D to larger scattering length, yields higher precision. However, we leave a careful exploration of the thermometric sensitivity in 1D to future work.

In order to understand the role of weak harmonic confinement, we also consider the case where the 1D gas is trapped by the harmonic potential $V_{\rm ext}(x) = \tfrac{1}{2}m\omega_0^2 x^2$. While in this case analytical solutions for the perturbed eigenfunctions are known~\cite{Busch1998}, we resort to numerical diagonalisation of $\hat{h}_0$ and $\hat{h}_1$ for simplicity. Again, reflection symmetry implies that only the even wavefunctions enter non-trivially into the determinant. The corresponding unperturbed energies are $E_n = \hbar\omega_0\left(2n+\tfrac{1}{2}\right)$ for $n=0,1,\ldots$. If $N_e$ atoms occupy even orbitals at $T=0$, the Fermi energy is thus defined by $E_F = \hbar\omega_0\left(2N_e-\tfrac{3}{2}\right)$, from which the Fermi wavevector $k_F = \sqrt{2mE_F}/\hbar$, temperature $T_F = E_F/k_B$ and time $\tau_F = \hbar/E_F$ can be derived. We hold $E_F$ constant, which is equivalent to keeping the density at the centre of the trap fixed. We also assume weak confinement, $\hbar\omega_0\ll E_F$, and focus on times less than the trap half-period, $\omega_0t< \pi$, in order to avoid partial recurrences~\cite{Sindona2013, Sindona2014}. Our numerical calculations follow the same truncation procedure described above.

The decoherence function for the harmonically trapped gas is compared to the homogeneous case in Fig.~\ref{fig:harmonic_1D}. We find that the norm of the decoherence function is very similar in both cases (main panel) but the phase of the decoherence function is noticeably different (inset). As a consequence, a temperature estimator based on the norm of the decoherence function may not need to account for details of the trap potential in the weakly confined regime. However, as discussed in the main text, at weak coupling the SLD becomes very sensitive to the phase of $v(t)$. The trap geometry must therefore be taken explicitly into account in order to achieve the highest precision. As shown in Fig.~\ref{fig:QSNR_1D}, the optimal precision attainable in the trapped gas is very similar to that of the homogeneous gas.